\documentclass[aps,prb, reprint, twocolumn, superscriptaddress, floatfix]{revtex4-2}

\usepackage{lipsum}  
\usepackage{graphicx}
\usepackage[cmyk,dvipsnames]{xcolor}
\usepackage{amsbsy}
\usepackage{amsmath}
\usepackage{amssymb}
\usepackage{amsfonts}
\usepackage{amsopn}
\usepackage{bigints}
\usepackage[utf8]{inputenc}
\usepackage[T1]{fontenc}
\usepackage{placeins}
\usepackage{float}
\usepackage{blindtext}
\usepackage{diagbox}
\usepackage[normalem]{ulem}
\usepackage{multirow}
\usepackage{bm}

\usepackage{hyperref}
\hypersetup{
    colorlinks,
    linkcolor={blue},
    citecolor={blue},
    urlcolor={blue}
}

\usepackage{algorithm}
\usepackage{algpseudocode}
\algnewcommand{\LineComment}[1]{\State // #1}
\algrenewcommand\textproc{\texttt}

\newcommand{\defcommand}[3][1]{
    \providecommand{#2}{}
    \renewcommand*{#2}[#1]{#3}
}

\defcommand[1]{\mat}
{
    \boldsymbol{#1}
}

\DeclareMathOperator{\tr}{tr}

\definecolor{pacificb}{HTML}{1CA9C9}

\newcommand{\fz}{Peter Gr\"unberg Institut and Institute for Advanced Simulation, Forschungszentrum J\"ulich and JARA, 52425 J\"ulich, Germany}
\newcommand{\iceland}{Science Institute and Faculty of Physical Sciences, University of Iceland, VR-III, 107 Reykjav\'{i}k, Iceland}
\newcommand{\rwth}{Department of Physics, RWTH Aachen University, 52056 Aachen, Germany}
\newcommand{\kalmar}{Department of Physics and Electrical Engineering, Linnaeus University, SE-39231 Kalmar, Sweden}

\begin{document}

\title{Identification of mechanisms of magnetic transitions using an efficient method for converging on first order saddle points}

\author{Hendrik Schrautzer}
\thanks{These authors contributed equally to this work.}
\affiliation{\iceland}

\author{Moritz Sallermann}
\thanks{These authors contributed equally to this work.}
\affiliation{\iceland}
\affiliation{\fz}
\affiliation{\rwth}

\author{Pavel F. Bessarab}
\email[Corresponding author: ]{pavel.bessarab@lnu.se}
\affiliation{\iceland}
\affiliation{\kalmar}

\author{Hannes J\'{o}nsson}
\affiliation{\iceland}

\begin{abstract}
    A method for locating first order saddle points on the energy surface of a magnetic system is described and several applications presented where the mechanism of various magnetic transitions is identified. The starting point for the iterative search algorithm can be anywhere, even close to a local energy minimum representing an initial state of the system, and the final state need not be specified. Convergence on a saddle point is obtained by inverting the component of the gradient along the minimum mode, thereby effectively transforming the neighbourhood of the saddle point to that of a local minimum. The method requires only the lowest two eigenvalues and corresponding eigenvectors of the Hessian of the system's energy and they are found using a quasi-Newton limited-memory Broyden–Fletcher–Goldfarb–Shanno solver for the minimization of the Rayleigh quotient without explicit evaluation of the Hessian. The method is applicable to large systems as the computational effort scales linearly with system size. Applications are presented to transitions in systems that reveal significant complexity of co-existing magnetic states, such as skyrmions, skyrmion bags, skyrmion tubes, chiral bobbers, and globules. When combined with rate theory within the harmonic approximation, the method can be used for simulations of the long timescale dynamics of complex magnetic systems characterized by multiple metastable states.
\end{abstract}




\maketitle

\section{Introduction}

The task of identifying the mechanism of possible transitions and estimating the corresponding rate within harmonic transition state theory (HTST)~\cite{wigner1938,vineyard1957} or Kramers/Langer theory~\cite{kramers1940,langer1969} involves finding low-lying first order saddle points (SPs) on the energy surface that specifies how the energy of the system varies as a function of the various degrees of freedom. A first-order SP is an extremal point, i.e. of zero gradient, where one and only one eigenvalue of the Hessian of the energy of the system is negative. It is challenging to locate a SP because of the need to maximize the energy with respect to one degree of freedom while it is minimized with respect to all others, as it is not known \textit{a priori} which degree of freedom to treat separately. If, in addition to the initial state, the final state of the transition is known, the minimum energy path (MEP) between the two corresponding minima on the energy surface can be found~\cite{jonsson_1998,bessarab2015}, and the relevant SP located as the point of highest energy along the path. However, various applications, such as a simulation of long timescale dynamics using the adaptive kinetic Monte Carlo algorithm~\cite{henkelman2001,jonsson2011}, necessitate the identification of likely transitions without any prior assumptions about final states. This is a more challenging task than the calculation of a minimum energy path for given initial and final states and can lead to the discovery of transition mechanisms into unexpected final states. SP searches can also be used as the basis for global optimization of an objective function in a broader context~\cite{plasencia2014}, as well as for path optimization~\cite{einarsdottir2012}, e.g. in calculations of tunneling within instanton theory~\cite{asgeirsson2018,vlasov2020} and radio wave propagation~\cite{nosikov2020}. An efficient algorithm for locating SPs can, therefore, have wide applicability.  

The development of SP search methods that do not require knowledge of the final state has, primarily, been in the context of atomic rearrangements, such as diffusion and chemical reactions~\cite{peters2017}. There, a SP search typically starts near the local energy minimum representing an initial state of the system and is carried out in two stages. Firstly, the system is driven out of the convex region where all eigenvalues of the Hessian of the energy are positive. Secondly, after the lowest eigenvalue turns negative, an algorithm for converging on the nearest SP is used. The first phase can be done in several different ways. In eigenvector following methods that are often used in studies of atomic rearrangements in small systems, the Hessian is evaluated and the eigenvalue problem solved to determine the eigenvectors. Various search paths are then generated starting from the minimum by following each one of the eigenvectors uphill until the convex region has been exited~\cite{peters2017}. A more efficient approach, especially for large systems, is to generate several starting points by displacing the system in some random way from the minimum and then moving uphill along the eigenvector corresponding to the lowest eigenvalue, the so-called minimum mode, until the lowest eigenvalue turns negative~\cite{henkelman1999}. Alternatively, the system can be pushed further along the direction defined by the difference between the starting configuration and the minimum while the energy is minimized in orthogonal directions~\cite{malek2000}. Several ways have been introduced for generating the starting configurations, for example Gaussian distributed displacements of the atoms from the minimum in some subregion of the system~\cite{pedersen2011} or evenly distributed points on the surface of a hypersphere~\cite{gutierrez2017}.

In the second stage, after escaping the convex region, the force acting on the atoms, i.e. the negative gradient of the energy with respect to atom coordinates, is modified by inverting the component along the minimum mode and the atoms displaced so as to zero this modified force. This ensures energy maximization along only the minimum mode and minimization along all other directions. Such a path eventually convergences on an SP. This technique is referred to as minimum mode following (MMF) method. Thus the search path is guided by the uphill direction of the minimum mode, while the energy is minimized with respect to all perpendicular directions. The minimum mode can be found by using, for example, the dimer~\cite{henkelman1999}, Lanczos~\cite{lanczos1950,malek2000} or Davidson~\cite{davidson1975,gutierrez2017} methods without even evaluating the Hessian matrix. 

Since the starting point of an SP search can be anywhere on the energy surface, MMF can also be used to converge on an SP starting from an approximate estimate, coming, for example, from a partially converged nudged elastic band calculation~\cite{jonsson_1998}.
Such a combination of an MEP finding method that is only partially converged followed by an SP search method starting from the point of highest energy along the obtained path, provides the most efficient way of converging on a SP corresponding to a transition to a given final state~\cite{asgeirsson2018}.

The rate of magnetic transitions, {\it i.e.} transitions where magnetic moments rotate, can in a similar manner be estimated by identifying SPs on the energy surface of a magnetic system~\cite{coffey2001,bessarab2012,bessarab2013}. When both initial and final states are specified, the minimum energy path can be found using the geodesic nudged elastic band method (GNEB) or the climbing image version, CI-GNEB where the highest energy image is pushed up to the maximum along the path~\cite{bessarab2015}. This method has been used, for example, to study the collapse of localized magnetic structures such as skyrmions~\cite{lobanov2016,cortes2017,vonmalottki2017,bessarab2018,heil2019,cortes2019,schrautzer2022,goerzen2023} as well as textures beyond skyrmions~\cite{hagemeister2018,desplat2019,kuchkin2020,kuchkin2023} including 3D states such as hopfions or skyrmion tubes~\cite{rybakov2015,muller2020,kuchkin2022,lobanov2023,sallermann2023,li2024}.

Similarly, the search for SPs without knowledge of the final state can be carried out for magnetic transitions. There, however, the curvature of the configuration space poses additional challenges as compared to the atomic rearrangements. The SP searches involve rotation of the magnetic moments using the force modified by an inversion along the minimum mode and projected onto the local tangent space of the current configuration. We refer to this as the \textbf{g}eodesic \textbf{m}inimum \textbf{m}ode \textbf{f}ollowing (GMMF) method.

Analogous to the MMF method for atomic rearrangements, the  GMMF method can be used to identify possible magnetic transitions from a given initial state without specifying a final state, thereby possibly discovering new and unexpected mechanisms and final states. Also, analogous to the atomic rearrangements, GMMF can be used in combination with a GNEB calculation that is not carried to completion, so as to converge on an SP corresponding to a transition to a known final state. The method can, furthermore, be generalized to find higher order saddle points, and this has been shown to be an efficient way to calculate excited electronic states~\cite{schmerwitz2023}.

A straightforward implementation of the GMMF method would involve the evaluation of the Hessian and calculation of at least the lowest eigenvalue and corresponding eigenvector, as was done by M\"uller {\it et al.}~\cite{muller2018}. However, the computational effort then increases rapidly with system size. The number of magnetic moments in relevant model systems can be large, often on the order of $10^5$ to $10^7$. Henceforth, even the mere evaluation of the full Hessian can require substantial effort. In particular, the explicit inclusion of long-range magnetostatic effects, as is frequently necessary in large three-dimensional systems, may not be feasible with such an approach. The reason is that the Hessian matrix then becomes dense and storing it can exceed the working memory capacity of a typical compute node, even for systems of moderate size. In addition, models that go beyond a simple Heisenberg approach, such as the non-collinear extension of the Alexander-Anderson model~\cite{bessarab2014,ivanov2020a} and density functional theory calculations, make the computation of second-order derivatives a significant task.

After the introduction of the GMMF method by M\"uller \textit{et al.}~\cite{muller2018}, it has been successfully used to identify mechanisms of magnetic transitions in several systems including the duplication of magnetic skyrmions~\cite{muller2018,muller2019}, transformation of defects in skyrmion lattices~\cite{bocquet2023}, annihilation of three-dimensional hopfions~\cite{sallermann2023}, and transitions in a two-dimensional dipolar spin glass~\cite{bocquet2023}. However, previous implementations of GMMF have been less than optimal in several respects, in that the full Hessian is explicitly evaluated and this limits the applicability of the method to rather small systems. Since the GMMF method strictly requires only the eigenvector corresponding to the lowest eigenvalue of the Hessian, the evaluation of the full Hessian is, in fact, not needed. A more efficient implementation of GMMF is, therefore, possible and is described in detail in this article. 

We report, in particular, an efficient way of computing the modified force in the GMMF method, using an iterative approach that does not require the evaluation of the Hessian. It is, thereby, applicable to large systems and more complex Hamiltonians than the Heisenberg form. Knowledge of the second lowest eigenvalue can also be useful to guide GMMF calculations, especially at mode crossings, as will be demonstrated below, and the method presented here is novel in that the lowest two eigenvalues and corresponding eigenvectors are found simultaneously, using the generalized Rayleigh quotient and an L-BFGS optimizer on the Grassmann manifold. 

The article is organized as follows. The method is described in Sec.~\ref{sec:method}. This is followed by Sec.~\ref{sec:model}, where the application systems are described. Results of several  GMMF calculations are presented Sec.~\ref{sec:results} featuring various magnetic structures in two- and three-dimensional systems and the performance is compared to that of a partial eigenvalue calculation with the Intel Math Kernel Library~\cite{intelMKL}. In the final section, Sec.~\ref{sec:conclusion}, the conclusions are summarized.
Appendix~\ref{app_sec:modecrossing_system} provides the setup of a system of four magnetic moments that is used as an example of mode crossings. Appendix~\ref{app:grassmann} provides the pseudocode for the optimization on the Grassmann manifold and Appendix~\ref{sec:app_eval_hess} explains the calculation of the full Hessian for reference. Appendix \ref{app:numerical_parameters} summarizes the numerical parameters used in the calculations.


\section{Method}
\label{sec:method}
One important aspect of the GMMF method is the correct consideration of the configuration space, $\mathcal{R}$, of magnetic systems which is a Riemannian manifold. Typically, the magnitude of the magnetic vectors is either assumed to be independent of orientation or treated as a fast variable within the adiabatic approximation. In the latter case, the magnitude is calculated for fixed orientations of the magnetic moments, which are considered to be slow variables~\cite{antropov1996}. In either case, the configuration space of a system with $N$ magnetic moments is a direct product of two-dimensional spheres $S_2$ associated with each magnetic moment vector: 
\begin{equation}
    \mathcal{R} = \bigotimes_N S_2 \subset \mathbb{R}^{3N},
    \label{eq:riemannian_manifold}
\end{equation}
giving rise to $n=2N$ degrees of freedom.
Displacements in the configuration space correspond to rotations of the magnetic moments. It is computationally advantageous to work in the embedding space $\mathbb{R}^{3N}$ and to eliminate the $N$ superfluous degrees of freedom by applying the projection operator approach~\cite{edelman1998,muller2018,varentcova2020,bocquet2023}. 

The iterative optimization procedure involved in the  GMMF method typically starts in the vicinity of the local energy minimum representing the initial state of the magnetic system. The climb up to the SP involves two stages: i) Escape from the convex region around the minimum, and ii) convergence on a first order SP on the energy surface. This two-stage procedure is to be repeated multiple times so as to identify, with some degree of confidence, all relevant SPs surrounding the given energy minimum. Convergence on various SPs is achieved by generating several different starting points near the energy minimum and/or by following different scenarios in the escape stage. For example, the convex region can be escaped by displacing the system from the initial state minimum along various eigenmodes of the system~\cite{muller2018}. In studies of atomic rearrangements, a method based on even distribution of points on a hypersphere in the configuration space has been used to sample different directions away from the initial state minimum~\cite{gutierrez2017,ohno2004,maeda2005}. Clearly, the choice of the escape strategy affects the efficiency of the method in finding as many distinct SPs as possible while keeping the number of the GMMF calculations to a minimum. Choosing the optimal escape strategy is an important task which, however, goes beyond the scope of the present study.
\begin{figure*}
    \centering
    \includegraphics{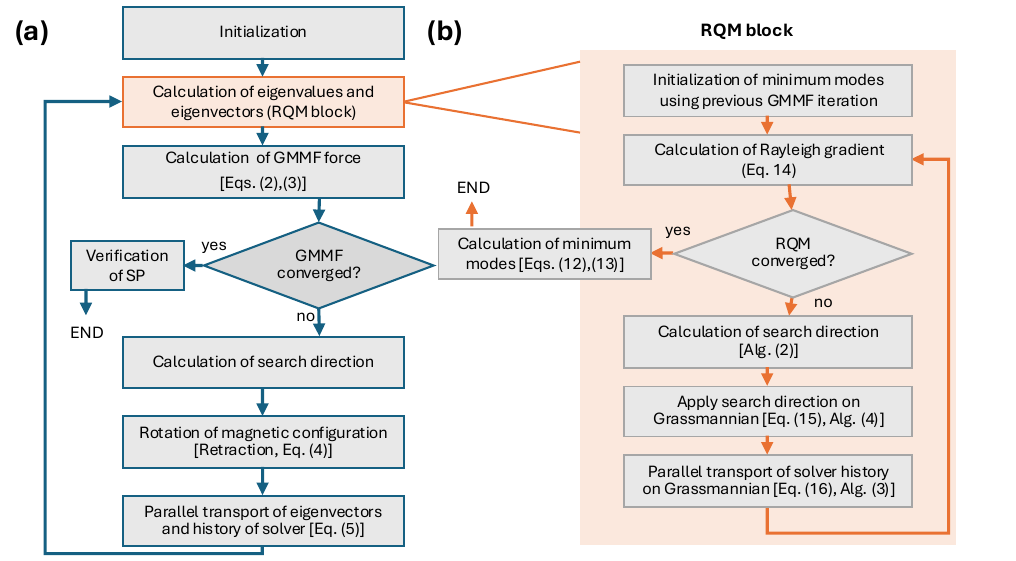}
    \caption{Flowchart for the algorithm of the GMMF method~(a) and for the RQM eigensolver method~(b) as presented in Sec.~\ref{sec:method}.  In each GMMF iteration, the RQM block is executed to calculate the eigenvalues and eigenvectors of the Hessian for the current magnetic configuration. The pseudocode for the algorithms describing the RQM method is given in App.~\ref{app:grassmann}}. 
    \label{fig:method_flowchart}
\end{figure*}
Here, the focus is on an efficient implementation of the second stage of the SP search -- the SP convergence stage. 
At a certain point during the escape stage, the minimum eigenvalue of the Hessian turns negative, which is an indication that a basin of attraction~\cite{henkelman1999} for an SP has been reached and that the SP convergence stage of the GMMF method can be started. The steps involved in the GMMF convergence stage are summarized in the flowchart in Fig.~\ref{fig:method_flowchart}(a).

\begin{figure}
    \centering
\includegraphics{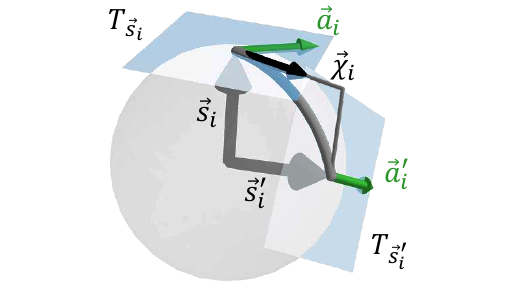}
    \caption{Visualization of rotating a single magnetic moment $\vec{s}_i$ in the direction $\vec{\chi}_i\in T_{\vec{s}_i}$ (Retraction). Also, the parallel transport of an exemplary vector $\vec{a}_i\in T_{\vec{s}_i}$ (green) towards the tangent space of the retracted magnetic moment $T_{\vec{s}_i'}$ is shown.}
    \label{fig:spin_retract_ptrans}
\end{figure}

In contrast to the escape stage, the strategy for advancing the system during the convergence stage once the system is outside the convex region can be made more systematic. The corresponding rotations of the magnetic moments are guided by a modified magnetic force designed so as to carry out a maximization of the energy along a certain direction -- the inversion mode -- and minimization along all orthogonal directions. Near an SP, the inversion mode necessarily is the minimum mode, i.e. the eigenvector of the Hessian corresponding to the lowest eigenvalue. However, near the boundary of the convex  region, it can be advantageous to choose the inversion mode to be the Hessian’s eigenvector corresponding to the second lowest eigenvalue or even higher eigenvalue. Such a formulation eventually results in a convergence on a first-order SP rather than a minimum. Specifically, the GMMF force on the $i^\text{th}$ magnetic moment in the system is defined as 
\begin{equation}
    \vec{f}_i=\vec{b}_i-(\vec{b}_i\cdot\vec{s}_i)\vec{s}_i,
    \label{eq:force_modified}
\end{equation}
where $\vec{s}_i$ is the unit vector in the direction of the $i^\text{th}$ magnetic moment and $\vec{b}_i$ is the effective field whose component along the inversion mode is reversed. The modified effective field $\vec{b}_i$ is defined by the following equation:
\begin{equation}
    \vec{b}_i=-\vec{\nabla}_i E+2(\bm{\nabla}E\cdot\bm{q})\vec{q}_{i}.
    \label{eq:b_mod}
\end{equation}
Here, $E=E(\bm{s})$ is the energy as a function of the magnetic configuration defined by vector $\bm{s}=(\vec{s}_1,\vec{s}_2,\ldots,\vec{s}_N)$, $\bm{q}=(\vec{q}_1,\vec{q}_2,\ldots,\vec{q}_N)$ is the unit vector representing the inversion mode, and $\bm{\nabla}=(\vec{\nabla}_1,\vec{\nabla}_2,\ldots,\vec{\nabla}_N)$, with $\vec{\nabla}_i\equiv \partial/\partial \vec{s}_i$. By construction, the GMMF force $\bm{f}=(\vec{f}_1,\vec{f}_2,\ldots,\vec{f}_N)$ is orthogonal to the magnetic moments, i.e., lies in the tangent space of $\mathcal{R}$ for a given magnetic configuration $\bm{s}$: $\bm{f}\in T_{\bm{s}}$ (see Fig.~\ref{fig:spin_retract_ptrans} for an illustration of the tangent space for a single magnetic moment).

The inversion along the minimum mode makes the effective field correspond to the neighborhood of a minimum of $E(\bm{s})$ rather than that of an SP. This is the basic idea of the method: a transformation of the problem of locating a first-order SP into the much simpler task of gradient-based minimization. In particular, the GMMF force is used to guide an iterative advancement of the magnetic structure toward an SP using some numerical optimization method, preferably one that accounts for the curvature of the configuration space $\mathcal{R}$ via retraction and parallel transport~\cite{absil2008book}. The concepts of retraction -- movement on a manifold in a certain direction without leaving the manifold -- and parallel transport -- translation of geometrical data on a manifold -- are implemented for magnetic systems via rotations and illustrated for a single magnetic moment in Fig.~\ref{fig:spin_retract_ptrans}. Specifically, retracting a magnetic moment $\vec{s}_i$ in the direction $\vec{\chi}_i\in T_{\vec{s}_i}$, with $T_{\vec{s}_i}$ being the corresponding tangent space, to the updated position $\vec{s}_i^{\text{ }'}$ characterized by the tangent space $T_{\vec{s}_i^{\text{ }'}}$, as well as the parallel transport of a vector $\vec{a}_i$ (\textit{e.g.} the gradient or search direction from previous iterations in optimization methods with memory) from $T_{\vec{s}_i}$ to $T_{\vec{s}_i^{\text{ }'}}$ can be computed using:
\begin{align}
    \vec{s}_i^{\text{ }'}&=D_{\vec{\chi}_i}(\varphi)\vec{s}_i\label{eq:retraction_spin}\\
    \vec{a}_i^{\text{ }'}&=D_{\vec{\chi}_i}(\varphi)\vec{a}_i.
    \label{eq:ptrans_spin}
\end{align}
Here, the $3\times 3$ rotation matrix $D_{\vec{\chi}_i}(\varphi)$ is given by the Rodriguez rotation formula:
\begin{equation}
\label{eq:rodriguez}
    D_{\vec{\chi}_i}(\varphi)=I+\sin\varphi_iK_{\vec{\chi}_i}+(1-\cos\varphi_i)K_{\vec{\chi}_i}^2,
\end{equation}
where $I$ is a $3\times 3$ identity matrix
and $\varphi_i=\varphi|\vec{\chi}_i|$ is the angle between $\vec{s}_i^{\text{ }'}$ and $\vec{s}_i$. 
The matrix $K_{\vec{\chi}_i}$ is given by:
\begin{equation}
    K_{\vec{\chi}_i}=\begin{pmatrix}0      & -k^z_i & k^y_i  \\
        k^z_i  & 0      & -k^x_i \\
        -k^y_i & k^x_i  & 0\end{pmatrix},
\end{equation}
with $k^x_i,k^y_i,k^z_i$ being the Cartesian components of the rotation axis $\vec{k}_i$:
\begin{equation}
    \vec{k}_i=\vec{s}_i\times \frac{\vec{\chi}_i}{|\vec{\chi}_i|}.
\end{equation}
The retraction and parallel transport on the configuration space $\mathcal{R}$ can be described by the direct sum of the $3$ by $3$ rotation matrices for the individual magnetic moments:
\begin{equation}
    \mathcal{D}_{\bm{\chi}}(\varphi)=
    \begin{pmatrix}D_{\vec{\chi}_1}(\varphi) & 0 & \dots & 0 \\ 0 & D_{\vec{\chi}_2}(\varphi) & \dots & 0 \\ \vdots & \vdots & \ddots & \vdots \\ 0 & 0 & \dots & D_{\vec{\chi}_N}(\varphi)\end{pmatrix}.
    \label{eq:rotation_directsum}
\end{equation}

While many different optimization algorithms can be used, we employ here a limited-memory Broyden-Fletcher-Goldfarb-Shanno (L-BFGS) solver~\cite{liu1989} adapted to magnetic systems~\cite{ivanov2021}. Typically, the calculation is considered to be converged when the magnitude of the GMMF force has dropped below some threshold value [see Fig.~\ref{fig:method_flowchart}~(a)]. After that, the candidate SP is verified by ensuring that one and only one eigenvalue of the Hessian is negative. Occasionally, following the GMMF force can lead back into the convex region around a minimum. This happens if the inversion mode becomes nearly orthogonal to the energy gradient. In this case, the GMMF optimization is essentially just an energy minimization with a  displacement along the negative energy gradient. Such re-entrance into the convex region can be detected when the lowest eigenvalue of the Hessian becomes positive, indicating the need to restart the SP search.

\begin{figure*}
    \centering
    \includegraphics{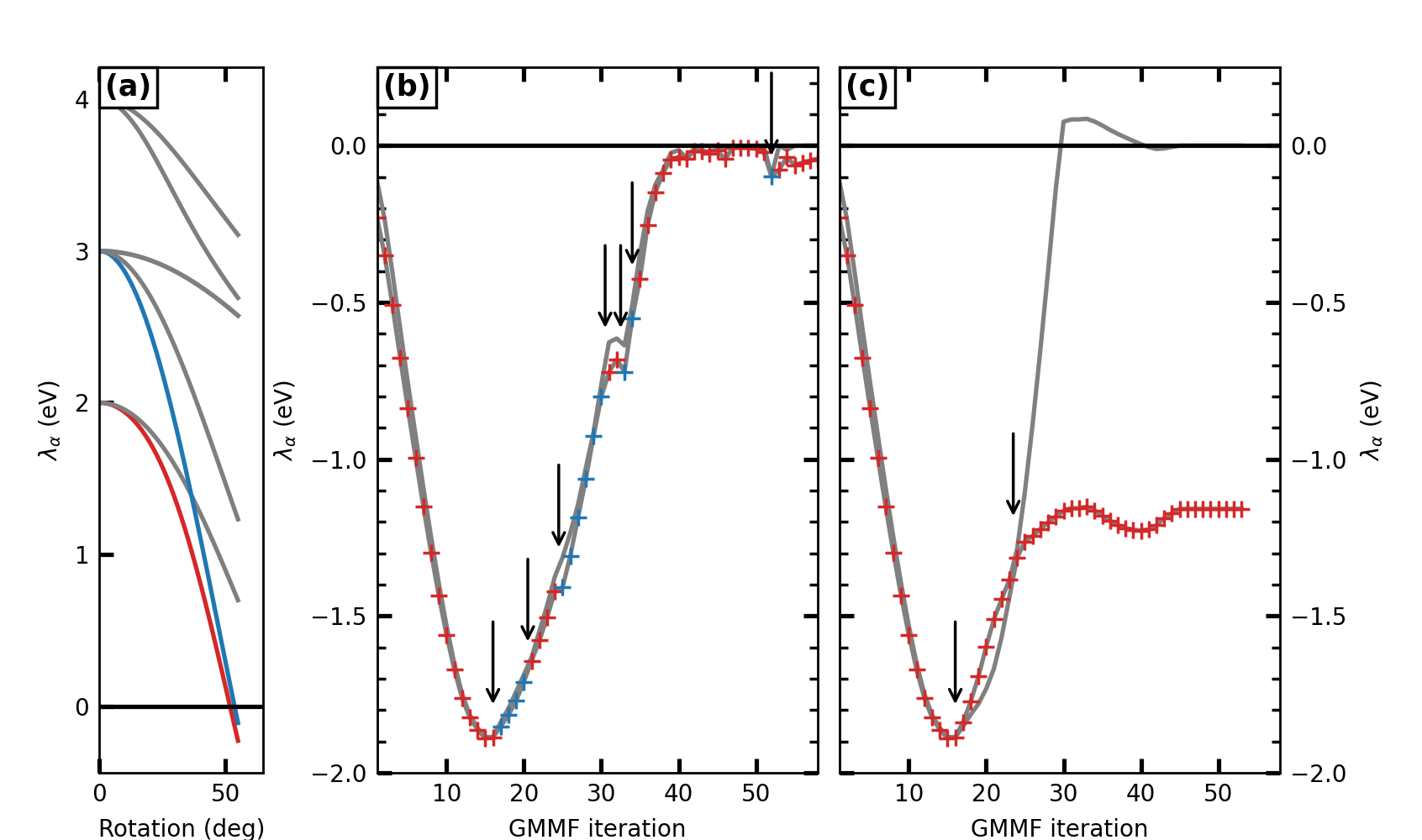}
    \caption{GMMF calculation applied to the system of four magnetic moments described in App.~\ref{app_sec:modecrossing_system}. (a):  The lowest eigenvalues of the Hessian as functions of the angle of rotation of two moments with respect to the initial parallel-aligned state during the escape stage. The red and blue curves correspond to the modes that experience mode crossings during the convergence stage. The maximum rotation angle corresponds to the starting configuration for the convergence stage of the GMMF calculation. The lowest eigenvalues are nearly equal at the maximum rotation angle. (b,c): Evolution of the two lowest eigenvalues during the convergence stage of the GMMF calculation, following two different strategies. The mode crossing events are marked with black arrows. The color of the crosses codes the inversion mode. In (b), the inversion mode is always chosen to be the minimum mode. In (c), the definition of the inversion mode changes when a mode crossing occurs. In particular, the inversion mode corresponds to the Hessian's eigenvector associated with the second lowest eigenvalue between the mode crossing events.
}
    \label{fig:method_mode_crossing}
\end{figure*}

Calculating the GMMF force involves reversing the component of the effective field along the inversion mode. Typically -- and always towards the end of the convergence stage --, the inversion along the minimum mode is carried out. However, during GMMF optimization, a crossing of the two lowest eigenvalues and corresponding modes may occur. Such a situation is depicted in Fig.~\ref{fig:method_mode_crossing}. The corresponding calculations were performed for a system with four magnetic moments described in App.~\ref{app_sec:modecrossing_system}. In such a case, the use of only the minimum mode for the inversion at each iteration step leads to an abrupt change in the GMMF force. Most of the times, this does not cause problems and the method converges on an SP after the mode crossing event. However, there are some situations where the two lowest modes are close to being degenerate and yet describe completely different changes in the magnetic structure. Then the use of only the minimum mode for the inversion may lead to repetitive mode crossing events [see Fig.~\ref{fig:method_mode_crossing}~(b)]. In this case, the consistency between iteration steps is lost due to multiple abrupt changes in the GMMF force, preventing the optimization procedure from converging. By selecting the eigenvector corresponding to the second-lowest eigenvalue as the direction of inversion for several iterations after the mode-crossing event can help prevent this unwanted behavior [see Fig.~\ref{fig:method_mode_crossing}~(c)]. By allowing for the inversion along several different modes can also increase the likelihood of identifying more distinct SPs. It is, therefore, useful to calculate the two lowest modes, not just the minimum mode. Furthermore, it is necessary to calculate the second lowest eigenvalue in order to verify that the candidate SP is of first order, i.e. only the lowest eigenvalue of the Hessian is negative.
 
The evaluation of the Hessian eigenvectors corresponding to the two lowest 
eigenvalues needs to be carried out efficiently as it tends to be a computational bottleneck of the GMMF method. Commonly used algorithms for finding the lowest few eigenvalues and corresponding eigenvectors include the Lanczos method~\cite{lanczos1950} and extensions thereof, such as the Davidson method~\cite{davidson1975} and other even more sophisticated Krylov subspace algorithms~\cite{sleijpen2000}. In this work, we present an alternative approach where the lowest eigenpairs of the Hessian are obtained via the direct minimization of the generalized Rayleigh quotient. The high efficiency of the method is achieved by avoiding the evaluation of the full Hessian explicitly and by using the L-BFGS solver, adapted to account for the curvature of the Grassmann manifold on which the objective function is defined. A detailed description of the method is provided in the following subsections and the corresponding pseudocode presented in Appendix~\ref{app:grassmann}.

\subsection{Calculation of the two lowest modes using generalized Rayleigh quotient minimization}
\label{ssec:finding_the_lowest_modes}
The two lowest modes, i.e. the eigenvectors corresponding to the two lowest eigenvalues of the Hessian, can be identified using the following Rayleigh quotient: 
\begin{equation}
    R(X) = \tr \left(X^T H X\right), 
   \label{eq:rayleigh_general}
\end{equation}
where $H$ is the $2N\times 2N$ Hessian matrix and $X$ is a $2N\times 2$ matrix whose columns are orthonormal vectors parametrizing all possible two-dimensional subspaces of the local tangent space of $\mathcal{R}$.
The subspace containing the two lowest modes is spanned by the columns of $X_\mathrm{min}$ corresponding to a minimum of $R(X)$~\cite{edelman1998}: 
\begin{equation}
X_\mathrm{min}\colon R(X_\mathrm{min})\rightarrow \min.
    \label{eq:minimum_of_RQ}
\end{equation}
The representation of the two lowest modes $\bm{v}_\alpha$, $\alpha=1,2$ in the embedding $3N$-dimensional Euclidean space can be obtained from $X_\mathrm{min}$ using:
\begin{equation}
    \bm{v}_\alpha = UX_\mathrm{min}\tilde{\bm{v}}_\alpha,
    \label{eq:RR_1}
\end{equation}
where $U$ is the projector onto the local tangent space of $\mathcal{R}$ and $\tilde{\bm{v}}_\alpha$ are the eigenvectors of the following $2 \times 2$ eigenvalue problem: 
\begin{equation}
    (X_\mathrm{min}^T H X_\mathrm{min}) \tilde{\bm{v}}_\alpha = \lambda_\alpha \tilde{\bm{v}}_\alpha \quad \alpha \in \{1,2\}.
    \label{eq:RR_2}
\end{equation}

The Rayleigh quotient minimization (RQM) needs to be performed at every iteration of the GMMF procedure.
To achieve this, we have implemented an efficient method based on the L-BFGS solver equipped with the gradient of $R(X)$ on the Grassmannian manifold~\cite{edelman1998}: 
\begin{equation}
    \nabla R(X) = 2 \left[H X - X(X^T H X)\right].
\label{eq:nablaR}
\end{equation}
The pseudocode for the GMMF/RQM algorithm is presented in Appendix~\ref{app:grassmann}. 
The RQM calculations are highly efficient and typically converge to $X_\mathrm{min}$ in just a few L-BFGS iterations, especially when the initial guess for $X_\mathrm{min}$ is taken from the solution of a previous GMMF iteration.
In Sec.~\ref{ssec:appl_skyrmion_collapses}, the performance of the GMMF/RQM approach is compared with that of an implementation of GMMF equipped with a state of the art extremal eigensolver based on a Krylov-Schur~\cite{stewart2002krylov} method as implemented in the Intel Math Kernel Library~\cite{intelMKL}(GMMF/KS).

It is important to realize that the domain of $R(X)$ represents a curved, Grassmann manifold $G_{2N,2}$. Therefore, an efficient RQM requires implementation of retraction and parallel transport that adhere to the geometric constraints of $G_{2N,2}$. Given a search direction $\Lambda$ for the current iteration $X$, the retraction to the next iteration $X'$ on $G_{2N,2}$ must ensure orthonormality of the columns of $X'$. Here, we make use of the exponential map, which is a special kind of retraction following geodesics. On the Grassmann manifold, the action of the exponential map can be found by the compact singular value decomposition $\Lambda=P\Sigma V$ of the tangent vector $\Lambda$~\cite{edelman1998} (see App.~\ref{app:grassmann}, Alg.~\ref{alg:displace}):
\begin{equation}
    X' = XV\cos(\Sigma)V^T+P\sin(\Sigma)V^T
    \label{eq:retraction_grassmann}
\end{equation}
The L-BFGS algorithm that we use for finding $X_\text{min}$ approximates the second derivative of the Rayleigh quotient and combines information of the gradients and the search directions from the past few iterations. These history vectors are anchored in the tangent spaces of their point of calculation on the curved manifold. Thus, in each iteration of the RQM a history vector $A$ has to be transported from the tangent space of $X$ to the tangent space of $X'$ using $A'=TA$ with~\cite{edelman1998}:
\begin{equation}
    T = \begin{pmatrix}
        X V & P
    \end{pmatrix}\begin{pmatrix}
        -\sin\Sigma \\ \cos \Sigma
    \end{pmatrix}P^T+(I-PP^T),
    \label{eq:transport_grassmann}
\end{equation}
where $I$ is the $2N\times 2N$ identity matrix. A computationally efficient multiplication with this parallel transport matrix is given in App.~\ref{app:grassmann}, Alg.~\ref{alg:transport}.

Finally, the global minimum of Eq.~\ref{eq:rayleigh_general} is given by $X_\text{min}$ which spans the same subspace as the sought-for two eigenvectors of the Hessian. Note that all $2N\times 2$ matrices whose orthonormal columns span the same subspace represent the same point on $G_{2N,2}$. Thus, the eigenvectors corresponding to the two lowest eigenvalues of the Hessian are linear combinations of the columns of $X_\text{min}$, and are determined by solving Eq.~(\ref{eq:RR_2}).

A method for computing the Hessian for magnetic systems is described in Refs.~\cite{muller2018} and Ref.~\cite{varentcova2020} and is also summarized in App.~\ref{sec:app_eval_hess}. Explicit evaluation of the Hessian is feasible for systems of moderate size and/or systems with only short-range interactions as the Hessian matrix then becomes sparse. The explicit Hessian is used in the KS implementation of the GMMF method.

The evaluation of the gradient of the Rayleigh quotient does not, however, require the explicit evaluation of the Hessian; only its action on $X$ is needed [see Eq. (\ref{eq:nablaR})]. The action of the Hessian on $X$ can be obtained through separate matrix-vector multiplications. These can be calculated using a finite difference scheme for the energy gradient, as described in the following subsection. This approach avoids the explicit evaluation of the Hessian, providing a significant speedup in the GMMF calculations.

\subsection{Action of the Hessian matrix}
\label{ssec:action_hessian_matrix}
\begin{figure}
    \centering
    \includegraphics[width=0.9\linewidth]{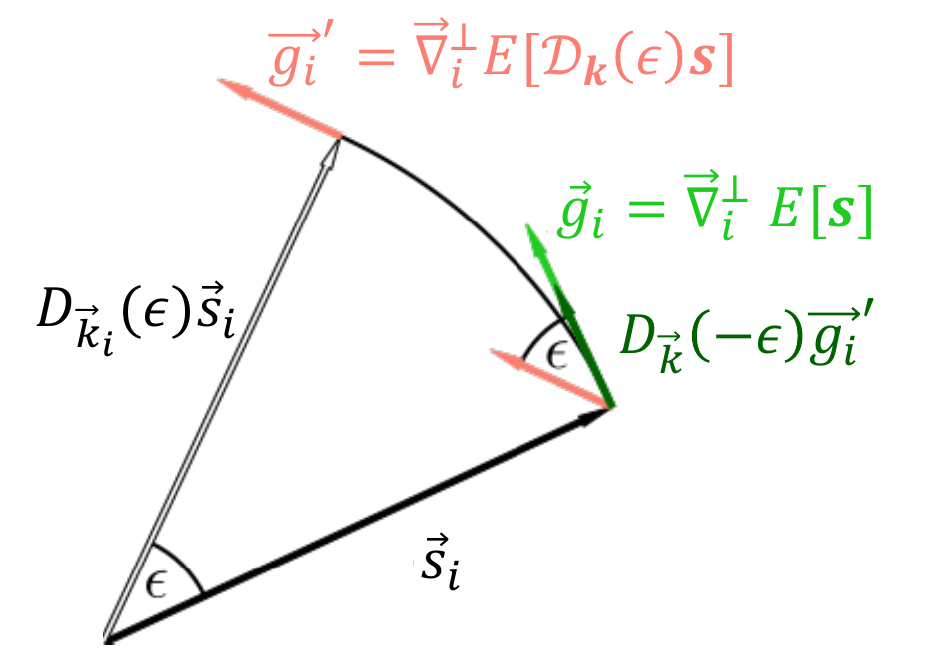}
    \caption{Illustration of the quantities involved in calculating the action of the Hessian 
    by forward finite difference
    approximation
    with a displacements parameter $\epsilon$ for a single magnetic moment $\Vec{s}_i$. To calculate the finite difference of the gradient $\vec{g}_i$ along a given direction, the magnetic moment is first displaced (rotated) along this direction and the gradient $\Vec{g}_i^{\text{ }'}$ is calculated for the rotated magnetic moment $D_{\vec{k}_i}(\epsilon)\vec{s}_i$. Parallel transport ($\mathcal{D}_{\hat{k}_i}(-\epsilon)\Vec{g}_i^{\text{ }'}$) is then applied towards the tangent space of the initial magnetic moment $\vec{s}_i$ to ensure that both gradients are defined within the same tangent space for the calculation of the finite difference.}
    \label{fig:finite_difference}
\end{figure}

Calculation of the matrix-vector multiplication $H\bm{x}$, where $\bm{x}$ is a column of $X$, is needed for the gradient-based minimization of the Rayleigh quotient [see Eq.~(\ref{eq:nablaR})] for $\nabla R(X)$. It can be performed without the explicit evaluation of the Hessian $H$ using a finite difference scheme for the energy gradient. In such a scheme, one needs to take into account the curvature of the configuration space. In particular, energy gradients for retracted magnetic configurations must be parallel-transported to the location of the current magnetic configuration so as to be anchored in the same tangent space [see Fig.~\ref{fig:finite_difference}]. The retraction and parallel transport are implemented through rotations [see Eqs.~(\ref{eq:retraction_spin}) and~(\ref{eq:ptrans_spin})] in the direction of $\bm{x}$, which can be expressed in terms of the vector $\bm{\chi}$ in the embedding $3N$-dimensional Euclidean space, thereby avoiding basis changes:
\begin{equation}
    \bm{\chi} = U\bm{x}.
\end{equation}
The forward finite difference scheme for computing $H\bm{x}$ is used:
\begin{equation}
    H\bm{x}\approx U^T \frac{\mathcal{D}_{{\bm{\chi}}}(-\epsilon)\bm{\nabla}^\perp E\left[\mathcal{D}_{{\bm{\chi}}}(\epsilon)\bm{s}\right]-\bm{\nabla}^\perp E\left[\bm{s}\right]}{\epsilon}.
    \label{eq:findiff}
\end{equation}
Here, $\bm{\nabla}^\perp=(\vec{\nabla}_1^\perp,\dots,\vec{\nabla}_N^\perp)$, with $\vec{\nabla}_i^\perp=\vec{\nabla}_i-\vec{s}_i(\vec{s}_i\cdot\vec{\nabla}_i)$, $\epsilon>0$ is the displacement parameter, and projector $U^T$ yields the $2N$-dimensional representation of the action of the Hessian matrix as required in the context of the RQM algorithm.
Note that the angle of rotation is not uniform across the entire system. Instead, it is defined individually for each magnetic moment based on the components of $\bm{\chi}$ [see Eqs.~(\ref{eq:rodriguez})-(\ref{eq:rotation_directsum})].
The optimal value of the displacement parameter $\epsilon$ yielding the minimum error of the finite-difference was determined using the Richardson extrapolation scheme~\cite{brezinski2010,richardson1911,richardson1927} with the initial $\epsilon$ value of $10^{-6}$. If necessary, the finite-difference scheme can be extended to incorporate additional stencil points. However, we find that the forward finite-difference scheme, combined with the Richardson extrapolation, is sufficient for the RQM calculations. 
 
\subsection{Further analysis}
\label{ssec:further_considerations}
The GMMF/RQM method offers various synergies with other approaches, as listed below. On the one hand, it is straightforward to generate the MEP once a first-order SP and the corresponding unstable mode have been determined. To achieve this, a steepest descent path is generated from the SP on both sides toward the stable states. During the minimization, the cumulative geodesic distance can be used to control how many points are generated to provide a good discrete representation of the MEP, similar to the path that would be obtained by the GNEB method if both initial and final state minima were known beforehand.

On the other hand, the highest energy image of a partially converged GNEB calculation can be used as a starting point for a GMMF calculation. In this way, the fast convergence of the GMMF method on a first order SP is used instead of the larger computational effort of converging all images in a GNEB simulation. 

Once an SP has been found, the pre-exponential factor in the rate constant for the corresponding transition can be obtained within the HTST approximation~\cite{bessarab2012}. By identifying all relevant, low energy SPs connecting to a given initial state, a Monte Carlo procedure can be used to select which transition will occur next in a long timescale evolution of the system, as has been done for atomic systems in the adaptive kinetic Monte Carlo method~\cite{henkelman2001} where the system recursively traverses from one energy minimum to another {\it via} the identified SPs. Such a long timescale simulation at a finite temperature, combined with systematic coarse graining of the energy landscape~\cite{pedersen2012,chill2014}, can be the basis for simulated annealing search for the global energy minimum representing the most stable state of the system.


\section{Model}
\label{sec:model}
The GMMF/RQM method described in Sec.~\ref{sec:method} can be applied to any system where the energy depends on the orientation of magnetic moments $\bm{s}$, such as atomistic spin models~\cite{eriksson2017}. One example is the extended Heisenberg model which has been extensively used to describe various topological spin textures such as magnetic skyrmions and other even more complicated magnetic states~\cite{kuchkin2020}. 

To demonstrate the GMMF/RQM method, we use the following Hamiltonian on square [see Sec.~\ref{ssec:appl_skyrmion_collapses}] and cubic lattice [see Sec.~\ref{ssec:appl_skyrmiontube}] models:
\begin{align}
    \begin{split}
        E = &- J \sum\limits_{i,j}\Vec{s}_i\cdot\Vec{s}_j - D\sum\limits_{i,j}\hat{d}_{ij}\cdot(\vec{s}_i\times\Vec{s}_j)\\
        &-\mu\sum\limits_{i}\Vec{B}_\text{ext}\cdot\vec{s}_i.
    \end{split}
    \label{eq:hamiltonian}
\end{align}
Nearest-neighbor exchange ($J$) is included as well as nearest neighbor Bloch-type Dzyaloshinskii-Moriya interaction (DMI) ($D$), with the unit DMI-vector $\hat{d}_{ij}$ oriented along the connection line of the interacting magnetic moments $i$ and $j$. The magnitude of the magnetic moments is set to $\mu=1\,\mu_B$, with $\mu_B$ being the Bohr magneton. An external magnetic field $\Vec{B}$ is applied in the $z$-direction. The 2D system studied in Sec.~\ref{ssec:appl_skyrmion_collapses} is similar to the one studied by \citet{muller2018}. The parameter values chosen for the 3D system investigated in Sec.~\ref{ssec:appl_skyrmiontube} are the same as those used by \citet{rybakov2015}, see Tab.~\ref{tab:modelparameters}.
\begin{table}[t]
\centering
\caption{Parameters of the extended Heisenberg model [Eq.~\ref{eq:hamiltonian}] used in the calculations. The size refers to the number of unit cells in the $x$, $y$ and $z$ directions, respectively. For the first application, a range in parameter values was explored.
}
\begin{ruledtabular}
\begin{tabular}{cccccc}
System & Size &Section & $J$ (meV)  & $D$ (meV) & $B$ (T) \\ \hline
Square & $50\times 50\times 1$& \ref{ssec:appl_skyrmion_collapses} & $1$&$[0.2,0.8]$& $[2.0,5.0]$\\
Cubic  & $30\times 30\times 30$& \ref{ssec:appl_skyrmiontube}& $1$ & $0.45$& $2.8$ \\
\end{tabular}
\end{ruledtabular}
\label{tab:modelparameters}
\end{table}


\section{Results}
\label{sec:results}
\subsection{Skyrmion collapse mechanisms}
\label{ssec:appl_skyrmion_collapses}
\begin{figure}
    \centering
    \includegraphics{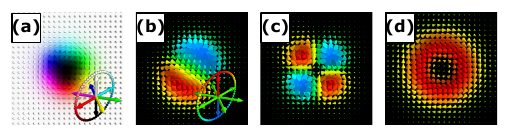}
    \caption{(a): Skyrmion corresponding to a local energy minimum on a square lattice with $J=1~\text{meV}$, $B=3.1~\text{T}$ and $D=0.55~\text{meV}$ [see Eq.~\ref{eq:hamiltonian}]. The inset explains the hue-saturation-lightness color scheme. While the azimuthal angle of the magnetic moments is represented by the hue, the lightness encodes the polar angle of the magnetic moments for a fixed saturation of $1.0$. (b-d): Eigenvectors of the Hessian corresponding to the translation mode (b), elongation mode (c), and breathing mode (d). The color codes the polar angle of the eigenvector components [See inset in (b)].}
    \label{fig:model_skyrmion_spectrum}
\end{figure}

\begin{figure}
    \centering
    \includegraphics{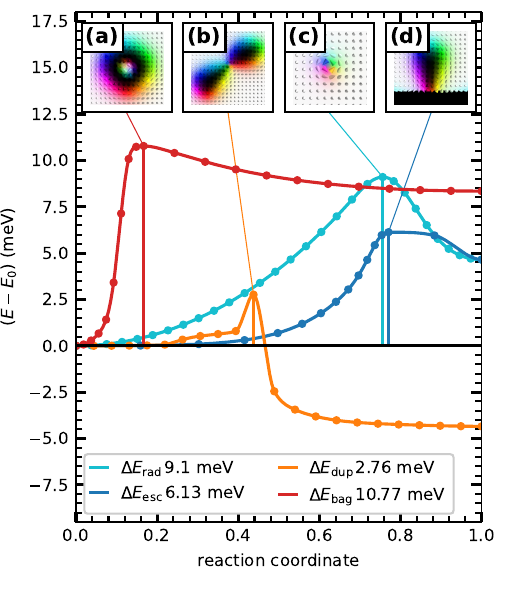}
    \caption{Minimum energy paths obtained by CI-GNEB calculations for skyrmions on a square lattice with $J=1~\text{meV}$, $B=3.1~\text{T}$ and $D=0.55~\text{meV}$ [see Eq.~\ref{eq:hamiltonian}]. The initial paths were generated using the SP configurations obtained with the GMMF algorithm [see Sec.~\ref{ssec:further_considerations}]. The insets show the first order SP configurations for the transition of the skyrmion into the skyrmion bag (a), the duplication of a skyrmion (b), the radially symmetric skyrmion collapse (c) and the escape of the skyrmion through the system boundary (d). The energy barriers defined as the energy difference between the SP state and the skyrmion state are represented by vertical lines, with their values provided in the legend.}
    \label{fig:results_skyrcollapse_paths}
\end{figure}

In this section, the application of the GMMF method to systems hosting skyrmions is demonstrated.
The energy-minimum skyrmion state [see Fig.~\ref{fig:model_skyrmion_spectrum}~(a)] serves as a starting configuration for the SP searches.
For $J=1$ meV, $D=0.55$ meV, and $B=3.1$ T, four distinct SPs corresponding to different skyrmion collapse mechanisms can be identified using the GMMF method:
\begin{enumerate}
    \item $\text{SP}_\text{rad}$ corresponding to a Bloch point-like configuration [Fig.~\ref{fig:results_skyrcollapse_paths}~(c)] on the path that connects the skyrmion state with the FM state via radially-symmetric shrinking of the skyrmion.
    \item $\text{SP}_\text{esc}$ characterizing the collapse of the skyrmion via an escape through the boundary of the system [Fig.~\ref{fig:results_skyrcollapse_paths}~(d)]. The SP configuration is a droplet-like deformed skyrmion attached to the system's boundary.
    \item $\text{SP}_\text{dup}$ corresponding to the splitting of a single skyrmion into two skyrmions (skyrmion duplication). The SP configuration is an eight-shaped deformed skyrmion [Fig.~\ref{fig:results_skyrcollapse_paths}~(b)].
    \item $\text{SP}_\text{bag}$ characterizing the transformation of the skyrmion state into the skyrmion bag state [Fig.~\ref{fig:results_skyrcollapse_paths}~(a)]. The SP configuration is a skyrmion with strongly canted magnetic moments at the skyrmion center.
\end{enumerate}
Each of these SPs corresponds to the configuration of highest energy of the MEP connecting the skyrmion state with the FM state, the skyrmin bag state or the state of two skyrmions. 
The MEPs are shown in Fig.~\ref{fig:results_skyrcollapse_paths}. 
They were obtained from CI-GNEB calculations where an initial path was generated by sliding down the energy surface from the SP on both sides toward the stable states [see Sec.~\ref{ssec:further_considerations}].
For the selected values of the Hamiltonian parameters, the energy of the ferromagnetic state is higher than the energy of the isolated skyrmion, and the energy of two skyrmions is the lowest.

To locate the four SPs, the convex region of the skyrmion energy minimum is escaped using four different strategies. Each strategy corresponds to a specific deformation of the magnetic configuration, thereby introducing tendencies that reflect the shape of the SP configurations. Three of the four deformations can be generated using the eigenvectors of the low curvature part of the eigenvalue spectrum of the skyrmion state. Those are shown in Fig.~\ref{fig:model_skyrmion_spectrum}~(b-d) and have been discussed in several publications~\cite{zhang2017,desplat2018,muller2018,vonmalottki2019,varentcova2020}. To reach SP$_\text{bag}$, a rotation of magnetic moments at the skyrmion center is generated.

\begin{figure}[t]
    \centering    
    \includegraphics{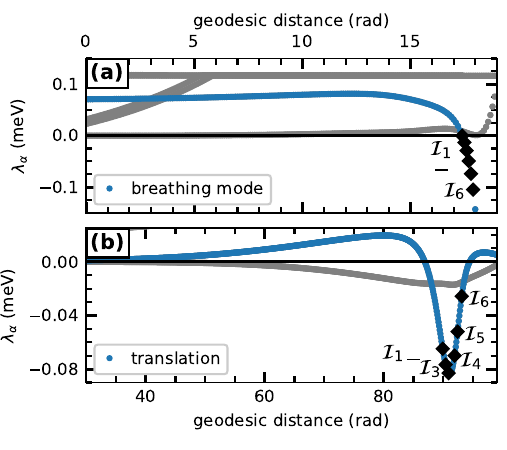}
    \caption{Evolution of the lowest eigenvalues of the Hessian during iterative displacement along the breathing mode (a) and the translation mode (b) starting from the skyrmion state. The cumulated geodesic distance along this displacement is shown on the $x$-axis. The eigenvalues of the minimum mode corresponding to six starting configurations $\mathcal{I}_1$ to $\mathcal{I}_6$ for subsequent GMMF simulations are marked with black diamonds. The parameters of the Hamiltonian are $J=1~\text{meV}$, $B=3.1~\text{T}$ and $D=0.55~\text{meV}$.}
    \label{fig:escape_radial_boundary}
\end{figure}

Iteratively displacing the skyrmion configuration along the eigenvector of the breathing mode [see Fig.~\ref{fig:model_skyrmion_spectrum}~(d)] shrinks or expands the skyrmion. Fig.~\ref{fig:escape_radial_boundary}~(a) shows how several of the lowest eigenvalues change as the skyrmion shrinks. At some point, one of the eigenvalues rapidly decreases and turns negative indicating that the convex region around the skyrmion state minimum has been escaped. A subsequent application of GMMF leads to $\text{SP}_\text{rad}$. A similar displacement along the eigenvector of one of the two degenerate translation modes [See Fig.~\ref{fig:model_skyrmion_spectrum}~(b)] moves the skyrmion towards the boundary of the system. Again, a sharp decrease of the lowest eigenvalue during this movement can be seen in Fig.~\ref{fig:escape_radial_boundary}~(b). However, the eigenvalue of the translation mode used for the displacements corresponds to the minimum mode only for a certain range of displacements [see Fig.~\ref{fig:escape_radial_boundary}~(b)]. The corresponding configurations mark suitable initial points for finding SP$_\text{esc}$ using GMMF.

\begin{figure}
    \centering
    \includegraphics{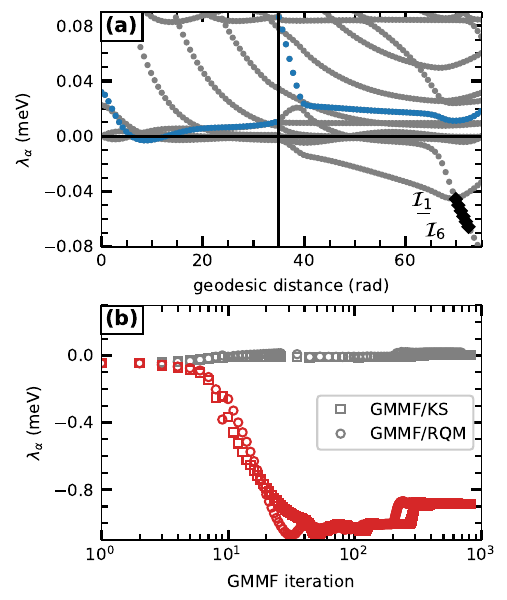}
    \caption{(a): The evolution of the lowest eigenvalues of the Hessian during the escape stage corresponding to iterative displacement of the system along the skyrmion elongation mode (before the black vertical line) and the mode describing a narrowing of the central part of the elongated skyrmion (after the black vertical line). Six suitable start configurations $\mathcal{I}_1-\mathcal{I}_6$ for the GMMF method are marked with black diamonds. (b): The evolution of the two lowest eigenvalues during the application of the GMMF/RQM and GMMF/KS method for the starting point $\mathcal{I}_1$. The final configuration of the GMMF algorithm corresponds to $\text{SP}_\text{dup}$. The parameters of the Hamiltonian are $J=1~\text{meV}$, $B=3.1~\text{T}$ and $D=0.55~\text{meV}$.}
    \label{fig:results_skyrcollapse_duplication}
\end{figure}

To create a tendency towards SP$_\text{dup}$ during the escape stage, an eigenvector of one of the two degenerate elongation modes [see Fig.~\ref{fig:model_skyrmion_spectrum}~(c)] is followed.
After a certain number of iterative displacements, a low-curvature mode emerges in the spectrum, which corresponds to a narrowing of the central part of the elongated skyrmion.
From this point onward [see the vertical black line in Fig~\ref{fig:results_skyrcollapse_duplication}~(a)], the iterative displacements are directed along the eigenvector associated with this mode.
The displacement along the narrowing mode is accompanied by a rapid drop in one of the eigenvalues. 
As soon as this mode becomes the minimum mode, a starting configuration for GMMF is chosen, and the subsequent GMMF simulation converges on $\text{SP}_\text{dup}$.

\begin{figure}
    \centering
    \includegraphics{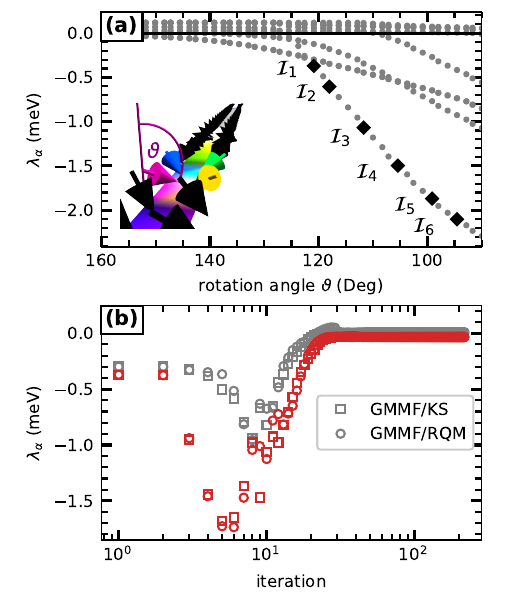}
    \caption{(a): The evolution of the lowest eigenvalues of the Hessian during the uniform rotation of the four magnetic moments in the core of the skyrmion. The angle between the magnetic moment and the $z$-axis is $\vartheta$, while the in-plane component of the magnetic moments always points towards the center of the skyrmion. Starting points for the GMMF algorithm are indicated by the black diamonds. (b): The evolution of the two lowest eigenvalues during the application of the GMMF/RQM and GMMF/KS method for the starting point $\mathcal{I}_1$. The final configuration of the GMMF algorithm corresponds to $\text{SP}_\text{bag}$. The values of the Hamiltonian parameters are $J=1~\text{meV}$, $B=3.1~\text{T}$ and $D=0.55~\text{meV}$.}
    \label{fig:results_skyrcollapse_skyrmionium}
\end{figure}

The escape strategy for $\text{SP}_\text{bag}$ is not based on the displacement along eigenmodes of the skyrmion, which is in contrast to other SPs. Instead, the magnetic moments at the skyrmion center are rotated toward the direction of the applied magnetic field. This is schematically shown in the inset in Fig.~\ref{fig:results_skyrcollapse_skyrmionium}~(a). During this deformation, one of the eigenvalues begins to decrease rapidly and eventually becomes the lowest eigenvalue. At this point, the application of the GMMF method leads to the identification of $\text{SP}_\text{bag}$.

The evolution of the two lowest eigenvalues during the application of the GMMF method equipped with the RQM eigensolver (GMMF/RQM) is shown in Fig.~\ref{fig:results_skyrcollapse_duplication}~(b) for $\text{SP}_\text{dup}$ and in Fig.~\ref{fig:results_skyrcollapse_skyrmionium}~(b) for $\text{SP}_\text{bag}$. For comparison, the same results are depicted for a GMMF method using the Krylov-Schur eigensolver [see Sec.~\ref{ssec:finding_the_lowest_modes}]. One can identify slight differences between the eigenvalues obtained in GMMF/RQM and GMMF/KS. This is expected due to the qualitative difference between the RQM and KS eigensolvers leading to slightly different search directions on the energy landscape. Accordingly, the eigenspectrum is not determined at exactly the same point in configuration space in the following iteration. Therefore, the GMMF method is stable against a slight variation of the optimization path, which is underlined by the behavior shown in Fig.~\ref{fig:results_skyrcollapse_duplication}~(b) and Fig.~\ref{fig:results_skyrcollapse_skyrmionium}~(b) as both methods converge reliably on the same SP. 

Figure~\ref{fig:results_skyrcollapse_skyrmionium}~(b) illustrates that the lowest and second-lowest eigenvalues can approach each other closely during GMMF calculations. This places a requirement on the eigensolver to be able to resolve spectra that are nearly degenerate, and the RQM method successfully meets this requirement in all situations encountered here.

In order
to get an impression of how the performance of the GMMF/RQM method compares to a straightforward implementation of the GMMF method equipped with a state-of-the-art eigensolver from the widely used Intel Math Kernel Library, we compare the wall times required by each implementation of the GMMF to achieve convergence on the four targeted SPs. For each SP, six different starting configurations ($\mathcal{I}_1,\dots,\mathcal{I}_6$) are used. The starting configurations, indicated by black diamonds in Figs.~\ref{fig:escape_radial_boundary}, \ref{fig:results_skyrcollapse_duplication}(a), and \ref{fig:results_skyrcollapse_skyrmionium}(a), differ in their distance from the convex region around the skyrmion state minimum. The wall time (averaged over 50 runs) and the number of iterations until convergence of the GMMF/RQM and GMMF/KS method for each starting configuration and SP are provided in Tab.~\ref{tab:GMMF_comp_effort}. The GMMF/RQM is on average ten times faster, although a slightly longer path is followed on the energy surface resulting in a small increase in the number of iterations. The calculations also demonstrate smaller sensitivity of the  GMMF method with respect to variation of the starting point. Comparison of the computational efficiency of the RQM and KS eigensolvers is not straightforward. The RQM method benefits significantly from using finite differences of the gradients instead of explicitly calculating the Hessian, even when the Hessian is sparse. Furthermore, it is not clear how the convergence behavior of an L-BFGS optimization of the Rayleigh Quotient compares to the KS method. A detailed analysis of this will be done in the future.

Note also that we do not restrict the application of the GMMF algorithm to regions of the configuration space where one and only one eigenvalue is negative. Such points can be hard to find especially in high-dimensional systems. Therefore, we allow for starting points where multiple eigenvalues are negative. As the energy is minimized along all directions orthogonal to the inversion mode, only one eigenvalue becomes negative at some point.

\begin{table*}[!htpb]
\centering
\caption{The mean wall time, $\tau$, and the number of iterations, $\kappa$, required to find SPs describing skyrmion transformations using the GMMF/RQM method, as ratios of those required using the GMMF/KS method.}
\begin{ruledtabular}
\begin{tabular}{ccccccccc}
Initial Point& \multicolumn{2}{c}{$\text{SP}_{\text{rad}}$} & \multicolumn{2}{c}{$\text{SP}_{\text{esc}}$}  & \multicolumn{2}{c}{$\text{SP}_{\text{dup}}$} & \multicolumn{2}{c}{$\text{SP}_{\text{bag}}$} \\ \hline
 & $\tau$ & $\kappa $ & $\tau$ & $\kappa $ & $\tau$ & $\kappa $ & $\tau$ & $\kappa $  \\
$\mathcal{I}_1$ & $0.14 \pm 0.04$ & $1.18$ & $0.08 \pm 0.05$& $1.50$ & $0.05\pm 0.02$ & $0.83$ & $0.12\pm 0.04$ & $1.73$\\
$\mathcal{I}_2$ & $0.14 \pm 0.04$ & $1.07$ & $0.10 \pm 0.05$& $2.05$ & $0.06\pm 0.03$ & $1.02$ & $0.13\pm 0.04$ & $1.84$\\
$\mathcal{I}_3$ & $0.14 \pm 0.05$ & $1.10$ & $0.10 \pm 0.04$& $1.93$ & $0.04\pm 0.02$ & $0.80$ & $0.14\pm 0.04$ & $1.89$\\
$\mathcal{I}_4$ & $0.14 \pm 0.05$ & $1.09$ & $0.11 \pm 0.04$& $2.09$ & $0.05\pm 0.03$ & $0.92$ & $0.10\pm 0.04$ & $1.22$\\
$\mathcal{I}_5$ & $0.14 \pm 0.04$ & $1.14$ & $0.10 \pm 0.04$& $1.97$ & $0.06\pm 0.04$ & $1.02$ & $0.11\pm 0.05$ & $1.45$\\
$\mathcal{I}_6$ & $0.13 \pm 0.05$ & $1.06$ & $0.06 \pm 0.04$& $1.09$ & $0.04\pm 0.03$ & $0.82$ & $0.11\pm 0.04$ & $1.42$\\
\end{tabular}
\end{ruledtabular}
\label{tab:GMMF_comp_effort}
\end{table*}

\begin{figure*}
    \centering
    \includegraphics{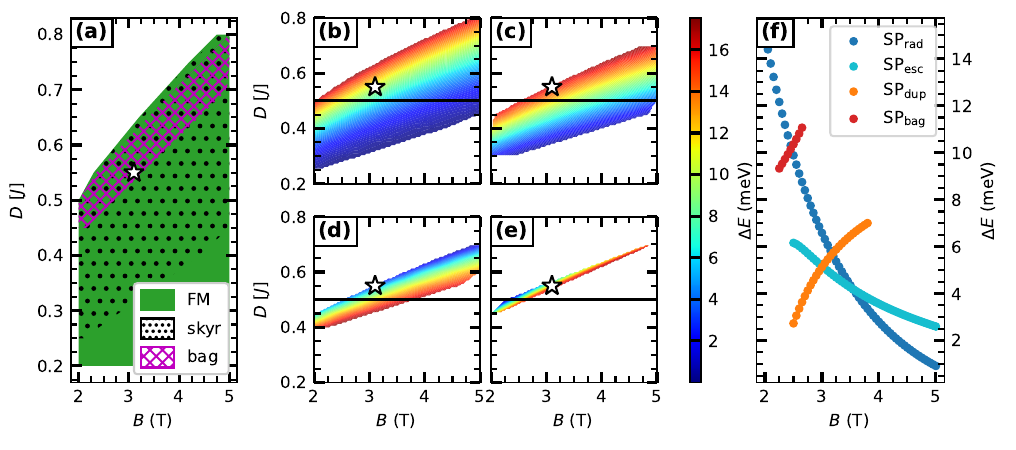}
    \caption{(a): Parameter domains of meta-stability of the ferromagnetic state (FM), the skyrmion (skyr) and the skyrmion bag (bag) obtained by energy minimization initiated with the corresponding states for the parameters $B=3.1$~T and $D=0.55$~meV (star symbol). The white area corresponds to the domain where all minimizations yield the spin spiral state. (b-e): Existence diagrams for the SPs $\text{SP}_\text{rad}$, $\text{SP}_\text{esc}$, $\text{SP}_\text{dup}$ and $\text{SP}_\text{bag}$ obtained via the GMMF/RQM method initialized with the corresponding SP at $B=3.1$~T and $D=0.55$~meV (star symbol). The colormap represents the energy barrier $\Delta E$ which is the difference of the energy of the SP and the energy of the skyrmion. (f): The energy barriers $\Delta E$ as functions of the applied magnetic field for a constant DMI parameter $D=0.55$~meV [horizontal lines in (b-e)].}
    \label{fig:results_spsampling_overview}
\end{figure*}

The efficiency of the GMMF/RQM method in locating SPs enables the straightforward identification of the domains in the parameter space where particular transition mechanisms exist, analogous to how magnetic phase diagrams are constructed by energy minimization. The domains of existence for $\text{SP}_\text{rad}$, $\text{SP}_\text{esc}$, $\text{SP}_\text{dup}$, $\text{SP}_\text{bag}$, along with those for the corresponding energy minima, are shown in Fig.~\ref{fig:results_spsampling_overview}. They were obtained by varying the $B$ and $D$ parameters and performing GMMF/RQM calculations for each point in the parameter space. Each calculation was initialized using the SPs determined for $B=3.1$~T and $D=0.55$~meV.

$\text{SP}_\text{rad}$ is present as long as the skyrmion state and the FM state co-exist [see Fig.~\ref{fig:results_spsampling_overview} (a),(b)]. Interestingly, the domains of existence for SPs do not always coincide with those of the adjacent energy minima. For example, the escape mechanism is not always available as the domain for $\text{SP}_\text{esc}$ is smaller than that for $\text{SP}_\text{rad}$ [see Fig.~\ref{fig:results_spsampling_overview} (b),(c)]. Similarly, $\text{SP}_\text{dup}$ is only present in the part of the skyrmion domain with large DMI [see Fig.~\ref{fig:results_spsampling_overview} (a),(d)], although duplication of a skyrmion is always possible through the emergence of a second skyrmion from the FM background via $\text{SP}_\text{rad}$. Finally, $\text{SP}_\text{bag}$ is limited to a very narrow region in the parameter space [see Fig.~\ref{fig:results_spsampling_overview} (e)].

The energy barrier for each skyrmion transformation can be obtained from the energy difference between the skyrmion state and the corresponding SP. Fig.~\ref{fig:results_spsampling_overview}~(f) shows how the energy barriers for each transition mechanism depend on the external magnetic field for a fixed DMI strength,  $D=0.5$~meV.

\subsection{Collapse mechanisms of 3D magnetic states}
\label{ssec:appl_skyrmiontube}
The advantage of using the finite difference scheme in the GMMF method as compared to evaluating the Hessian matrix explicitly, both in terms of computational time and memory requirements, becomes particularly clear when dealing with large systems. This is demonstrated by an application of the GMMF/RQM method to a cubic lattice with $30 \times 30 \times 30$ magnetic moments (see Tab.~\ref{tab:modelparameters} for the parameter values). Rybakov~\textit{et al.} have previously performed GNEB calculations for this system and demonstrated the transition of a skyrmion tube (SkT) to the helical phase state via two intermediate minima -- two chiral bobbers and a single chiral bobber (ChB)~\cite{rybakov2015}. Here, we demonstrate how the GMMF/RQM method can locate SPs without having to generate an initial path as is required in GNEB calculations. When investigating transitions between topological magnetic textures the generation of an appropriate initial paths is a challenging task and often requires intuition. For example, a linear interpolation between the initial and final states is often not sufficient for the identification of transitions involving translations of a magnetic texture. In the following, we demonstrate that the GMMF/RQM method is capable of locating the SPs without having to specify a final state and generate an initial path.

\begin{figure}
    \centering
    \includegraphics{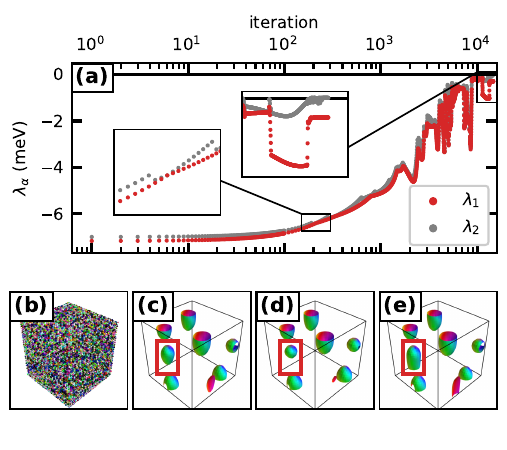}
    \caption{Application of the GMMF/RQM method to a cubic system with open boundaries in the $z$-direction initialized with a random configuration of the magnetic moments. The values of the Hamiltonian parameters can be found in Tab.~\ref{tab:modelparameters}. (a): Evolution of the two lowest eigenvalues $\lambda_1$, $\lambda_2$ during the GMMF/RQM calculation. The insets highlight a mode crossing as well as the convergence to a state with a single negative eigenvalue. The energy minimum configurations adjacent to the SP configuration (c) are shown in (d) and (e). Here, the magnetic configurations are visualized by the isosurfaces of $s_z=0$ where the color codes the in-plane component of the magnetization. A red outline highlights the transition from a globule (d) to a chiral bobber (e).}
    \label{fig:results_trumm3d_example}
\end{figure}

We first initialize the system with completely random magnetic orientations [See Fig.~\ref{fig:results_trumm3d_example}~(b)] and apply the GMMF method. Since the initial state of the system is far from an energy minimum, a negative eigenvalue already exists and the escape phase of an SP search is not needed. Although multiple negative eigenvalues of the Hessian are present at the beginning of the optimization, the GMMF method brings the system to a valley characterized by a single negative eigenvalue and eventually converges on a first-order SP corresponding to the magnetic configuration shown in Fig.~\ref{fig:results_trumm3d_example}~(c). Note that the optimization process involves multiple mode-crossing events, which correspond to the cusps in the curves shown in Fig.\ref{fig:results_trumm3d_example}(a), depicting the evolution of the two lowest eigenvalues during the GMMF calculation. The Hessian's eigenvector corresponding to the lowest eigenvalue is always chosen for the direction of inversion in this case.

The metastable configurations of the energy minima adjacent to the SP are shown in Fig.~\ref{fig:results_trumm3d_example}~(d,e). They involve combinations of various three-dimensional topological magnetic textures, such as chiral bobbers~\cite{rybakov2015} and globules~\cite{muller2020}. The SP configuration corresponds to the transformation of a globule to a chiral bobber. 

\begin{figure*}
    \centering
    \includegraphics{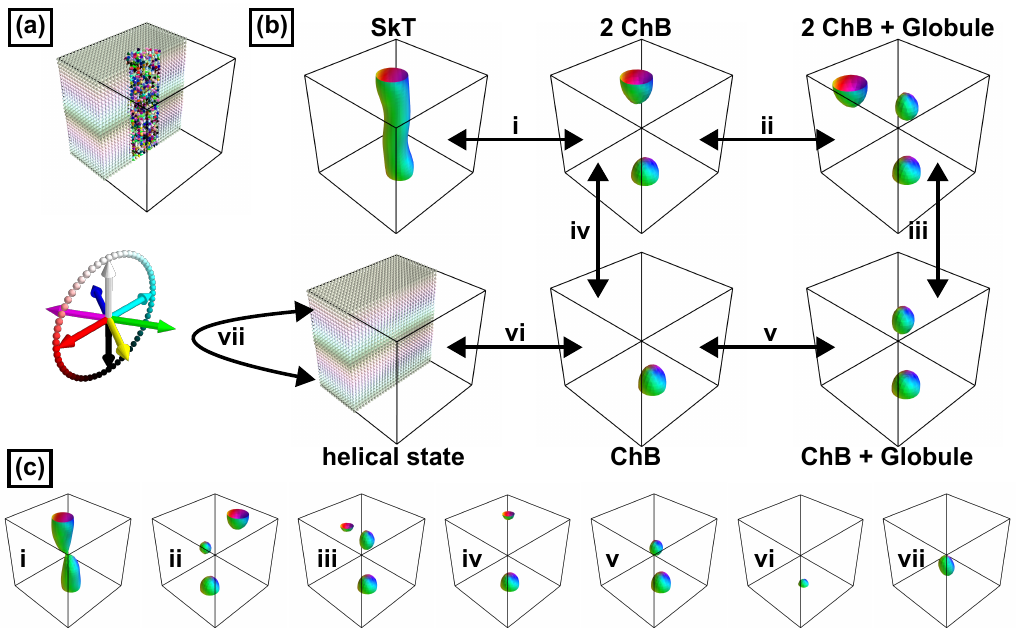}
    \caption{Magnetic configurations corresponding to local energy minima (b) and first-order SPs (c) obtained by the application of the GMMF/RQM method to the initial state shown in~(a). For each magnetic configuration, the iso-surface $s_z=0.0$ is shown, where the color codes the in-plane component of the magnetization. (b): Visualization of the local energy minimum configurations corresponding to the skyrmion tube (SkT), the helical state, and combinations of chiral bobbers (ChB) and globules. The arrows indicate transitions between the respective states and (c) visualizes the corresponding first-order SP configurations.}
    \label{fig:results_skt_map}
\end{figure*}

Initialization of the system in a completely random state typically leads to the formation of multiple localized magnetic states, which complicates a systematic investigation of various SPs and energy minima configurations in this system. Therefore, we choose a different initial configuration for the GMMF method: 
a cylinder with randomly oriented magnetic moments [see Fig.~\ref{fig:results_skt_map}~(a)] embedded in the helical phase, which corresponds to the ground state of the system for the chosen parameter values~\cite{rybakov2015}.
The application of the GMMF method to $500$ such random states resulted in the SP configurations shown in Fig.~\ref{fig:results_skt_map}~(c)
\footnote{The GMMF method also identifies additional SPs corresponding to translations of a magnetic configuration by less than a lattice constant. However, these SPs arise due to lattice effects and are not discussed further here.}. 
By calculating the energy minima adjacent to each SP by direct energy minimization, a graph showing possible transitions between the meta-stable configurations can be obtained, see Fig.~\ref{fig:results_skt_map}~(b).

The transitions (i, iv, vi) correspond to the decay of a skyrmion tube (SkT) into the helical phase via intermediate states corresponding to two and one chiral bobber (ChB), in agreement with the previous GNEB calculations in Ref.~\cite{rybakov2015}. M\"uller {\it et al.} have pointed out that isolated globules do not represent meta-stable configurations~\cite{muller2020}. However, it is possible to make them meta-stable near chiral bobbers, as transitions ii and v show. To our knowledge, it is a new finding that the isolated globule state can represent a first-order SP in the helical phase (vii).

By calculating the energy difference of the SPs and the adjacent energy minima $A$ and $B$, the energy barriers $\Delta E_{A\to B}$ and $\Delta E_{B\to A}$ can be obtained (See Tab.~\ref{tab:barriers_3d}). Note that the graph of states shown in Fig.~\ref{fig:results_skt_map}~(b) is likely not exhaustive. However, it demonstrates that many different states can be identified without relying on intuition. 

\begin{table*}
    \centering
    \caption{Energy barriers corresponding to the transitions between the local energy minima configurations shown in Fig.~\ref{fig:results_skt_map}(b).}
    \begin{ruledtabular}
        \begin{tabular}{cccccccc}
                   \multirow{2}{*}{\diagbox{State $A$}{State $B$}}                        & \multicolumn{6}{c}{Energy barrier $\Delta E_{A\to B}$(meV)}                                                                      \\
             & SkT                                                         & ChB    & 2 ChB   & ChB + Globule & 2 ChB + Globule & helical state \\ \hline
            SkT                            & $-$                                                         & $-$    & $23.13$ & $-$           & $-$             & $-$           \\
            ChB                            & $-$                                                         & $-$    & $17.14$ & $36.32$       & $-$             & $16.98$       \\
            2 ChB                          & $25.05$                                                     & $6.95$ & $-$     & $-$           & $31.84$         & $-$           \\
            ChB + Globule                  & $-$                                                         & $0.03$ & $-$     & $-$           & $16.09$         & $-$           \\
            2 ChB + Globule                & $-$                                                         & $-$    & $0.18$  & $13.31$       & $-$             & $-$           \\
            helical state                  & $-$                                                         & $7.34$ & $-$     & $-$           & $-$             & $42.02$       \\
        \end{tabular}
    \end{ruledtabular}
    \label{tab:barriers_3d}
\end{table*}


\section{Conclusions and discussion}
\label{sec:conclusion}

In the GMMF/RQM method presented here for the identification of first-order SPs on the energy surface of magnetic systems, convergence on SPs is achieved by maximizing the energy along the direction of one of the two lowest modes, {\it i.e.} the eigenvectors of the Hessian corresponding to the two lowest eigenvalues, and minimizing the energy along all other directions in the configuration space. Instead of the computationally demanding task of repeatedly evaluating the Hessian and solving the eigenvalue problem, we introduce an efficient methodology based on the direct minimization of the generalized Rayleigh quotient using only the energy gradients. This becomes particularly important for large systems and/or systems with dense Hessians. 

An important aspect of our methodology is the correct treatment of the geometry of the manifolds on which the optimization of the magnetic structure and minimization of the Rayleigh quotient are performed. In particular, optimization algorithms used in our method take into account the
curvature of the corresponding manifolds via retraction and parallel transport. The GMMF/RQM method exhibits a high degree of computational efficiency and stability which we demonstrate in applications to 2D and 3D magnetic systems. 

We show consistency with previously published results by identifying the three SPs corresponding to the mechanisms of skyrmion transformation reported in Ref. \citet{muller2018}. In addition, we identify the fourth mechanism corresponding to the transformation of the skyrmion into a skyrmion bag. 

An efficient computational scheme based on using only the energy gradients to calculate the action of the Hessian has made it possible to apply the GMMF method to a 3D system hosting chiral bobbers, skyrmion tubes, and globules. We obtain the collapse mechanism of a skyrmion tube via two intermediate states (chiral bobbers), in agreement with Ref. \citet{rybakov2015}. Moreover, we identify SPs corresponding to transitions between chiral bobbers and globule states. We show that the globule state, which cannot exist in isolation as a metastable state \cite{muller2020}, can represent a first-order SP in the helical phase. This information would not be accessible through standard energy minimization alone. It is important to realize that these results are obtained via straightforward application of the GMMF/RQM method without relying on subjective assumptions about the system.

While the GMMF/RQM method can be used independently to study transitions in various magnetic systems, it shows great promise when combined with other techniques such as the GNEB method or the CI-GNEB method where a climbing image is included to better converge on the SP~\cite{bessarab2015}. Instead of converging all images representing the path to a low tolerance, the calculation can be terminated at an earlier stage and the highest energy image used as a starting point for an efficient GMMF/RQM convergence on the SP. This is found 
to be the optimal strategy for magnetic systems, analogous to what has been found for atomic rearrangements~\cite{asgeirsson2021}.

The GMMF/RQM method can be integrated into a global optimization framework for magnetic textures by recursively traversing between local energy minima through first-order SPs~\cite{plasencia2014}. Within this framework, calculating transition rates on the fly -- using, for example, harmonic transition state theory -- would enable the simulation of long-timescale dynamics through the adaptive kinetic Monte Carlo algorithm~\cite{henkelman2001,jonsson2011}.

To summarize, the GMMF/RQM method represents a valuable tool for studying transitions in various magnetic systems. This becomes particularly important in the context of topological magnetism where various 2D and 3D states have recently been discovered.

Possible extensions of the method include the inversion of the magnetic force components along several modes simultaneously in order to locate higher order SPs~\cite{schmerwitz2023}. In this respect, the presented RQM methodology can easily be extended to the calculation of more modes.

The first phase of the SP search, namely the escape from the convex region, remains a challenging task. In this respect, organization of the GMMF escape stage using the symmetries of the magnetic textures appeaars to be promising. This is a subject of future research.

\begin{acknowledgments}
    M.S. is grateful to A. Goswami and G. P. M\"uller for fruitful discussions.
    H.S. is grateful to M. G\"orzen and S. Heinze for fruitful discussions. This work received financial support from the Icelandic Research Fund (grants No. 239435 and No. 217750), the University of Iceland Research Fund (grants No. 15673 and 15661), the Swedish Research Council (grant No. 2020-05110), and the Crafoord Foundation (grant No. 20231063). The calculations were carried out at the high-performance computing resources available at the Kiel University Computing Centre and at the Icelandic Research e-Infrastructure facility supported by the Icelandic Infrastructure Fund.
\end{acknowledgments}
M.S. and H.S. contributed equally to this work.
\appendix

\section{System of four magnetic moments illustrating mode crossing}
\label{app_sec:modecrossing_system}
To illustrate the mode crossings in Fig.~\ref{fig:method_mode_crossing}, a system of four magnetic moments placed in the corners of a square is considered.
Only the nearest neighbor moments are coupled via the exchange $J$ and a magnetic field in the negative $z$-direction is applied.
Furthermore, the uniaxial anisotropy favours alignement along the $\pm z$-direction.
The Hamiltonian of the system reads:
\begin{align}
    \begin{split}
    E = &- J(\Vec{s}_1\cdot\Vec{s}_2+\Vec{s}_1\cdot\Vec{s}_3+\Vec{s}_2\cdot\Vec{s}_4+\Vec{s}_3\cdot\Vec{s}_4)\\
    &-\mu\sum\limits_{i=1}^{4}\Vec{B}_\text{ext}\cdot\vec{s}_i - \sum\limits_{i=1}^{4}K (\vec{s}_i\cdot\vec{e}_z)^2
    \end{split}
    \label{eq:hamiltonian_supp}
\end{align}
with $J=0.5$~eV, K=$1.0$~eV and $\Vec{B}_\text{ext}=-1.0\hat{e}_z$~T as well as $\mu=1~\mu_B$.
For these parameters, the parallel alignment of the magnetic moments in the $+z$-direction is a local energy minimum configuration.
From this minimum configuration, a suitable start configuration for the application of the GMMF method is obtained by rotating two magnetic moments sitting on opposite corners of the square. 
The lowest eigenvalues spectrum of the system during this rotation is visualized in Fig.~\ref{fig:method_mode_crossing}~(a).

\section{Pseudocode for optimization of the Rayleigh quotient}
\label{app:grassmann}
The optimization is represented by the four Algorithms. The main routine is given in Alg.~\ref{alg:lbfgs_grassmann}. To compute the search direction, we use the well known "two loop recursion", listed in Alg.~\ref{alg:two_loop_recursion}.
Alg.~\ref{alg:lbfgs_grassmann}, in combination with Alg.~\ref{alg:two_loop_recursion}, constitutes a fairly standard L-BFGS method with some exceptions: A minor one is that the algorithm is being applied to matrices instead of -- as perhaps more commonly seen -- vectors.
The other, much more important, is the use of the \texttt{Parallel transport} and \texttt{Retraction} functions to move across the Grassmannian manifold, in a way that respects its inherent geometry. 
These two are implemented in Alg.~\ref{alg:transport} and Alg.~\ref{alg:displace}, respectively, and make use of equations given by Edelmann \emph{et. al.} \cite{edelman1998}.

\begin{figure}[!ht]
\begin{algorithm}[H]
    \caption{L-BFGS optimizer for the generalized Rayleigh Quotient}
    \label{alg:lbfgs_grassmann}
    \begin{algorithmic} 
        \Require Initial point $X$ on the Grassmann manifold $G_{2,2N}$, Hessian matrix $H$, tolerance $\epsilon$, maximum memory $m$
        \Ensure Solution $X$ that spans the minimal invariant subspace of $H$
        \State Initialize $k \gets 0$
        \State Initialize the history $S \gets []$, $Y \gets []$, $\rho \gets []$

        \While{$k < \text{maxIterations}$}
            \LineComment{Here, we denote the action of the Hessian as $HX$}
            \LineComment{In practice, we use the finite difference scheme,}
            \LineComment{described in App.~\ref{app:numerical_parameters}}
            \State $G_k \gets 2X_k \left[ HX_k - X_k (X_k^T H X_k) \right]$
            \If{$\|G_k\|_F < \epsilon$}
                \State \textbf{break}
            \EndIf
            \LineComment{Transport the history to the current tangent frame}
            \If{$\text{len}(S) > 0$}
                \State $S_i \gets \mathtt{Transport}(X_{k-1}, S_i, P, \Sigma, V) \quad \forall S_i \in S$
                \State $Y_i \gets \mathtt{Transport}(X_{k-1}, Y_i, P, \Sigma, V) \quad \forall Y_i \in Y$
            \EndIf
            \If{k > 1}
                \LineComment{Append to the history}
                \State Append $\Lambda$ to $S$
                \State Append $G_{k} - \mathtt{Transport}(G_{k-1}, P, \Sigma, V)$ to $Y$
                \State Append $\tr\left[(X_{k} - X_{k-1})^{T}(G_{k} - G_{k-1}) \right]^{-1}$ to $\rho$
                \If{ $\text{len}(S) > m$}
                    \State Delete oldest entry of $S, Y$ and $\rho$
                \EndIf
                \LineComment{Curvature rejection condition}
                \If{$\rho_{-1} < 0$}
                    \LineComment{Reset the history}
                    \State $S \gets []$, $Y \gets []$, $\rho \gets []$
                \EndIf
            \EndIf
            \LineComment{Compute the search direction via two loop recursion}
            \State $\Lambda \gets \mathtt{TwoLoopRecursion}(S, Y, \rho, G_k)$
            \LineComment{Perform compact singular value}
            \LineComment{decomposition of $\Lambda$ such that $Q = P \Sigma V^T$}
            \State $P, \Sigma, V \gets \mathtt{SVD}(\Lambda)$
            \LineComment{Update $X$ via parallel transport}
            \State $X \gets \mathtt{Displace}(X,P,\Sigma,V)$
            \State $k \gets k + 1$
        \EndWhile
    \end{algorithmic}
\end{algorithm}
\end{figure}

\begin{figure}[!ht]
\begin{algorithm}[H]
    \caption{Two loop recursion}
    \label{alg:two_loop_recursion}
    \begin{algorithmic}
        \Function{\texttt{TwoLoopRecursion}}{$S$, $Y$, $\rho$, $G$}
            \State $\Lambda \gets G$
            \For{$i = \text{len}(S)$ downto $1$}
                \State $\alpha_i \gets  \rho_i \text{tr}(S_i^T Q)$
                \State $\Lambda \gets \Lambda - \alpha_i Y_i$
            \EndFor
            \If{$\text{len}(S) > 0$}
                \LineComment{Initial diagonal approximation}
                \LineComment{for the inverse Hessian}
                \State $H_0^k \gets \rho_i^{-1} \sqrt{\tr \left( {Y_k}^T Y_k\right)}$ 
                \State $\Lambda \gets H_0^k \Lambda$
            \EndIf
            \For{$i = 1$ to $\text{len}(S)$}
                \State $\beta \gets \rho_i \text{tr}(Y_i^T \Lambda)$
                \State $\Lambda \gets \Lambda + (\alpha_i - \beta) S_i$
            \EndFor
            \LineComment{Minus sign for minimization}
            \State \Return $-\Lambda$
        \EndFunction
    \end{algorithmic}
\end{algorithm}
\end{figure}

\begin{figure}[!ht]
\begin{algorithm}[H]
    \caption{Parallel transport}
    \label{alg:transport}
    \begin{algorithmic} 
        \Function{Transport}{$X$, $A$, $P$, $\Sigma$, $V$}
           \LineComment{The parentheses are necessary to ensure a }
           \LineComment{computationally efficient order of evaluation}
           \State $T_1A \gets -X V \sin(\Sigma) (P^T A)$
           \State $T_2A \gets P \cos(\Sigma) (P^T A)$
           \State $T_3A \gets A - P (P^T A)$
           \State \Return $T_1A + T_2A + T_3A$
        \EndFunction
    \end{algorithmic}
\end{algorithm}
\end{figure}

\begin{figure}[t]
\begin{algorithm}[H]
    \caption{Retraction}
    \label{alg:displace}
    \begin{algorithmic} 
        \Function{\texttt{Retraction}}{$X$, $P$, $\Sigma$, $V$}
            \State $X' \gets X V \cos(\Sigma) V^T + P\sin(\Sigma)V^T$
            \LineComment{Prevent accumulation of numerical errors}
            \State Orthonormalize columns of $X'$ with QR algorithm
            \State \Return $X'$
        \EndFunction
    \end{algorithmic}
\end{algorithm}
\end{figure}

\section{Evaluation of the Hessian Matrix}
\label{sec:app_eval_hess}
The Hessian matrix for a magnetic system characterized by $N$ constraints on the length of the magnetic moments can be computed as follows.
First, the $3N\times3N$ matrix $H^{3N}$ of the second order derivatives of Eq.~\eqref{eq:hamiltonian} with respect to the Cartesian components of the magnetic moments is calculated.
After that, the shape operator $\mathcal{L}$ is subtracted from $H^{3N}$. 
The shape operator is a diagonal matrix of the following form
\begin{equation}
    \mathcal{L} = 
    \begin{pmatrix}
        L_1 & \hdots & 0\\
        \vdots & \ddots & \vdots\\
        0 & \hdots & L_N
    \end{pmatrix},
\end{equation}
with 
\begin{equation}
    L_i = \begin{pmatrix}
        \vec{s}_i \cdot \vec{\nabla}_iE & 0 &0\\
        0 & \vec{s}_i \cdot \vec{\nabla}_iE & 0\\
        0 & 0 & \vec{s}_i \cdot \vec{\nabla}_iE
    \end{pmatrix}.
\end{equation}
Finally, the matrix $U\in\mathbb{R}^{3N\times 2N}$ is applied, that projects the embedding space vectors into the tangent space of the current configuration $\bm{s}$ [see Ref.~\cite{muller2018, varentcova2020}]:
\begin{align}
    H=U^T(H^{3N}-\mathcal{L})U.
    \label{eq:2Nhessian}
\end{align}

\section{Numerical parameters for GMMF calculations}
\label{app:numerical_parameters}
The GMMF calculations are considered converged if the maximum norm of the effective field components drops below a value of $10^{-12}$~eV:
\begin{align}
    \underset{i\in \{1,\dots,N\}}{\operatorname{max}}|\vec{b}_i|<10^{-12}~\text{eV}
    \label{eq:conv_crit_TRUMM}
\end{align}
The L-BFGS based GMMF algorithm is used without a line search procedure. As described in Ref.~\cite{ivanov2021} we restrict the maximum steplength with a parameter $\vartheta_\text{max}$.
The number of memory quantities taken into account is $m=3$ for all calculations in this publication. For the GMMF calculations presented in Sec.~\ref{ssec:appl_skyrmion_collapses} we always chose $\vartheta_\text{max}^\text{GMMF}=0.5$. Also the L-BFGS solver of the RQM method was set to $\vartheta_\text{max}^\text{RQM}=0.5$. The calculation of the two lowest modes within the RQM method is considered converged if the Frobenius norm of the Rayleigh gradient drops below $10^{-8}$ eV.

\FloatBarrier

\bibliographystyle{apsrev4-2}
\bibliography{bibliography}
\end{document}